\documentclass[english]{article}
\usepackage[T1]{fontenc}
\usepackage[utf8]{inputenc}
\usepackage{geometry}
\usepackage{color}
\usepackage{babel}
\usepackage{float}
\usepackage{amsmath}
\usepackage{amsthm}
\usepackage{graphicx}
\usepackage{setspace}
\usepackage[unicode=true,pdfusetitle,
 bookmarks=true,bookmarksnumbered=false,bookmarksopen=false,
 breaklinks=false,pdfborder={0 0 1},backref=false,colorlinks=false]
 {hyperref}

\makeatletter
\theoremstyle{definition}
\newtheorem{defn}{\protect\definitionname}
\theoremstyle{plain}
\newtheorem{assumption}{\protect\assumptionname}
\theoremstyle{definition}
\newtheorem{problem}{\protect\problemname}
\theoremstyle{plain}
\newtheorem{criterion}{\protect\criterionname}
\theoremstyle{definition}
\newtheorem{sol}{\protect\solutionname}
\theoremstyle{plain}
\newtheorem{lyxalgorithm}{\protect\algorithmname}
\theoremstyle{remark}
\newtheorem{rem}{\protect\remarkname}
\theoremstyle{plain}
\newtheorem{lem}{\protect\lemmaname}

\makeatother

\providecommand{\algorithmname}{Algorithm}
\providecommand{\assumptionname}{Assumption}
\providecommand{\criterionname}{Criterion}
\providecommand{\definitionname}{Definition}
\providecommand{\lemmaname}{Lemma}
\providecommand{\problemname}{Problem}
\providecommand{\remarkname}{Remark}
\providecommand{\solutionname}{Solution}

\begin{document}
\begin{doublespace}
\begin{center}
\textbf{\textcolor{black}{\Large{}Imitation in the Imitation Game}}{\Large\par}
\par\end{center}

\begin{center}
\textbf{Ravi Kashyap }
\par\end{center}

\begin{center}
\textbf{SolBridge International School of Business / City University
of Hong Kong}
\par\end{center}

\begin{center}
October 02, 2017
\par\end{center}

\begin{center}
\par\end{center}

\begin{center}
of Psychological, Emotional, Social, and Cognitive Factors on Decision
Making in Financial Markets
\par\end{center}

\begin{center}
AMS: 91G10 Portfolio theory; 68Q32 Artificial intelligence; 68T05
Learning and adaptive systems
\par\end{center}

\begin{center}
\tableofcontents{}\newpage{}
\par\end{center}
\end{doublespace}
\begin{doublespace}

\section{Abstract}
\end{doublespace}

\begin{doublespace}
\noindent We discuss the objectives of automation equipped with non-trivial
decision making, or creating artificial intelligence, in the financial
markets and provide a possible alternative. Intelligence might be
an unintended consequence of curiosity left to roam free, best exemplified
by a frolicking infant. For this unintentional yet welcome aftereffect
to set in a foundational list of guiding principles needs to be present.
A consideration of these requirements allows us to propose a test
of intelligence for trading programs, on the lines of the Turing Test,
long the benchmark for intelligent machines. We discuss the application
of this methodology to the dilemma in finance, which is whether, when
and how much to Buy, Sell or Hold.
\end{doublespace}
\begin{doublespace}

\section{\label{sec:The-Circle-of}The Circle of Investment}
\end{doublespace}

\begin{doublespace}
\noindent On the surface, it would seem that there is a repetitive
nature to portfolio management, making it highly amenable to automation.
But we need to remind ourselves that the reiterations happen, under
the purview of a special kind of uncertainty, that applies to the
social sciences. (Kashyap 2016) goes into greater depth on how the
accuracy of predictions and the popularity of generalizations might
be inversely related in the social sciences. In the practice of investment
management and also to aid other business decisions, more data sources
are being created, collected and used along with increasing automation
and with attempts to bring greater intelligence in the decision process
(Tambe 2014; Bughin 2016; Provost \& Fawcett 2013).

\noindent \textbf{\textit{Artificial Intelligence (AI) in finance
is a topic that has gained increased attention and priority among
market participants. This article has been written to be of general
interest, since it paints a clear picture and illuminates the basic
ideas and techniques, required to make the incubation of intelligence
a reality (sufficient technical details are provided but have been
relegated to the appendix and other references). The interrelationship
of different approaches and the central questions that remain open
are also discussed. We consider how automation can be useful for the
financial services industry and specifically articulate points regarding
the what / how / why of applying AI to investment management.}}

\noindent It is important to bear in mind that due to the game theoretic
nature of the financial markets, specifically with regards to investing
and trading, improvements in data sources and related technology will
only be beneficial, if they are better than competing players (Gupta
\& George 2016; Barney 1995 discuss why investments alone do not generate
competitive advantage and instead firms need to create capabilities
that rival firms find hard to match). (Fama 1970) is a discussion
of fair games and efficient markets; (Kyle 1985; Foster \& Viswanathan
1990) solve for the Nash equilibrium, (Osborne \& Rubinstein 1994),
when trading is viewed as a game between market makers and traders.
(Kashyap 2016) has a proof of trading costs as a zero sum game. For
different types of zero sum games and methods of solving them, see:
(Von Neumann \& Morgenstern 1953; Laraki \& Solan 2005; Hamadène 2006).
(Bodie \& Taggart 1978; Bell \& Cover 1980; Turnbull 1987; Hill 2006;
Chirinko \& Wilson 2008) consider zero sum games in the financial
context.

\noindent If Alice and Red Queen of the Wonderland fame (Carroll 1865;
1871; End-note \ref{enu:The-Red-Queen's}) were to visit Hedge-Fund-Land
(or even Business-Land), the following modification of their popular
conversation would aptly describe the situation today, ``My dear,
here we must process as much data as we can, just to stay in business.
And if you wish to make a profit you must process at-least twice as
much data.'' 

\noindent We could also apply this to HFT-Land (HFT, High Frequency
Trading: Biais \& Woolley 2011; Menkveld 2013; Brogaard, Hendershott
\& Riordan 2014; Cartea, Jaimungal \& Ricci 2014; End-note \ref{enu:HFT})
and say: “My dear, here we must trade as fast as we can, just to stay
in business. And if you wish to make a profit, you must trade at-least
twice as fast as that.”, while reminiscing that the jury is still
out on whether HFT is Good, Bad or Just Ugly and Unimportant (Budish,
Cramton \& Shim 2015 argue that the high-frequency trading arms race
is a symptom of flawed market design and that financial exchanges
should use frequent batch auctions instead of the currently predominant
continuous limit order book; Brogaard, Hendershott \& Riordan 2017
highlight that some HFT activity could be harmful for liquidity; in
contrast, Li, Cooper \& Van Vliet 2017 indicate that high-frequency
trading has a liquidity provision effect and improves the execution
quality of low-frequency orders; Chaboud, Chiquoine, Hjalmarsson\&
Vega 2014 suggest that algorithmic trading causes an improvement in
the price efficiency of foreign exchange markets. Kirilenko, Kyle,
Samadi \& Tuzun 2017 study the events of May 6, 2010, that became
known as the Flash Crash and show that HFTs did not cause the Flash
Crash, but contributed to it by demanding immediacy ahead of other
market participants). 

\noindent A consensus (perhaps the only one) among most participants
seems to be that automated trading dominates the markets, though there
seems to be a significant amount of debate on the specific problems
that require better decision making\textcolor{black}{{} (}for more HFT
pros and cons see: Vuorenmaa 2013; Savani 2012\textcolor{black}{;
Hagströmer \& Norden 2013); (for the ethical angle see: Davis, Kumiega
\& Van Vliet 2013; Angel \& McCabe 2013; Cooper, Davis \& Van Vliet
2016). Efforts at using learning techniques and artificial neural
networks, (}Haykin 2004\textcolor{black}{), is not new to finance
(}Hawley, Johnson \& Raina 1990\textcolor{black}{; Kryzanowski, Galler
\& Wright 1993; Wong \& Selvi 1998; }Bahrammirzaee 2010\textcolor{black}{;
Cartea, Jaimungal \& Kinzebulatov 2016); but the limited success should
inform us to learn from the mistakes and try to make some fundamental
alternations to our approach and perspective. }

\noindent \textcolor{black}{Any attempt at creating artificial financial
agents can benefit immensely from the knowledge that has been accumulated
using models in which the participants are not fully rational and
their imperfect decisions have helped to explain financial phenomenon,
which is the field of behavioral finance (Barberis \& Thaler 2003)}.
Kumiega \& Van Vliet 2012 consider the behavioral aspects of algorithmic
trading; (Wang, Keller \& Siegrist 2011) show using surveys on risk
perceptions of investment products that respondents perceived easier-to-understand
products as less risky, which was likely driven by the familiarity
bias (Huberman 2001 provides compelling evidence that people invest
in the familiar while often ignoring the principles of portfolio theory;
also see: Fox \& Levav 2000; Seiler, Seiler, Harrison \& Lane 2013).
(Brown \& Cliff 2005; Shu \& Chang 2015; Au, Chan, Wang \& Vertinsky
2003; Chung, Hung \& Yeh 2012) discuss the influence of investor sentiment
on financial markets.

\noindent \textbf{\textit{\textcolor{black}{This background of the
financial landscape, prompts us to not focus on improved decision
making specific to narrow problems but to uncover the general principles
that might be necessary for increased intelligence. Once we outline
these foundation principles, we provide individual testable hypothesis,
though we need to remind ourselves that greater the coherence between
the components, better the intended outcome.}}}

\noindent \textbf{\textit{To the best of our knowledge, this is the
first known instance of a modified Turing Test for trading or investment
decision-making (section \ref{subsec:Acing-the-Turing}). We also
discuss an improvement of Searle's Chinese Room (a key argument in
AI) that is applicable for portfolio management(section \ref{subsec:Mexican-Chihuahua-solving}).
The financial services industry is among the leading investors in
AI; hence the answers / clues provided in this paper to the problems
of buying and selling assets can be immensely applicable outside finance,
to a wide cross-section of the business community.}}
\end{doublespace}
\begin{doublespace}

\section{A Profitable Benchmark for Brainpower}
\end{doublespace}

\begin{doublespace}
\noindent The problem of automation, with non-trivial decision making,
or designing intelligence artificially for application in the financial
markets, can be a rather trivial task, depending on which trader's
brainpower acts as our gold standard. As a first step, we recognize
that one possible categorization of different fields can be done by
the set of questions a particular field attempts to answer. Since
we are only the creators of different disciplines, but not the creators
of the world in which these fields need to operate, the answers to
the questions posed by any domain can come from anywhere, or, from
phenomenon studied under a combination of many other disciplines.

\noindent Hence, the answers to the questions posed under the realm
of automated trading (AT) or AI, can come from seemingly diverse subjects,
such as, physics, biology, mathematics, chemistry, marketing, finance,
psychology, economics, music, theater and so on. As we embark on the
journey to apply the knowledge from other fields to automated portfolio
management, (APM), we need to be aware that APM is ``Simply Too Complex'',
since all of time, portfolio management has just been about beating
a benchmark (and many times, this benchmark is all about taking profits
and avoiding losses, easier defined than done, as any portfolio manager
or trader would reckon). The complications are mainly to select the
right standards to compete with. To facilitate a reference point,
for the rest of the article, we define automated portfolio management
as below.
\end{doublespace}
\begin{defn}
\begin{doublespace}
\noindent \textit{Automated Portfolio Management (APM) or Artificial
Intelligence in Portfolio Management (AIPM) is the ability to connect
elements of previously attained information to effect a portfolio
management decision. Nothing lasts forever and hence, no decision
is good forever; but the longer a decision serves its purpose, the
greater the intelligence involved in making that decision.}
\end{doublespace}
\end{defn}
\begin{doublespace}
\noindent With this definition, it should become clear that AIPM requires
the ability to collect pieces of information and to connect them towards
a decision making goal. Since decisions are not going to be valid
indefinitely, we need to continue to use these abilities to effect
later decisions or improve upon decisions already made. We use the
term agent below to refer to our creations, which are expected to
display intelligence. (Russell \& Norvig 2016) is a comprehensive
discussion of the concept of an intelligent agent. (Wooldridge \&
Jennings 1995) discuss the most important theoretical and practical
issues associated with the design and construction of intelligent
agents. (Hand 1998) makes a case for the tools of statistics in the
arms race for collecting and mining data towards the goal of better
decision making.
\end{doublespace}
\begin{doublespace}

\subsection{\label{subsec:Automation-for-What}Automation for What Sake?}
\end{doublespace}

\begin{doublespace}
\noindent To be precise, this is not about automation to barter for
the Japanese drink (though, that seems like a wise exchange, in terms
of financial value, and might have been attempted many times before).
Below, we make a case for why automation is not only unavoidable,
but necessary for modern portfolio management. We adapt certain core
concepts from (Kashyap 2017), which considers the goal of creating
intelligence for the purpose of intelligence alone, and tailor it
to the nuances of dealing with uncertainty in the financial markets.

\noindent A central aspect of our lives is uncertainty and our struggle
to overcome it. Over the years, it seems that we have found ways to
understand the uncertainty in the natural world by postulating numerous
physical laws. The majority of the predictions in the physical world
hold under a fairly robust set of circumstances and cannot be influenced
by the person making the observation, and they stay unaffected if
more people become aware of such a possibility. In the social sciences,
the situation is exactly the contrary. (Popper 2002) gave a critique
and warned of the dangers of historical prediction in social systems.
A hall mark of the social sciences is the lack of objectivity. Here
we assert that objectivity is with respect to comparisons done by
different participants and that a comparison is a precursor to a decision.
\end{doublespace}
\begin{assumption}
\begin{doublespace}
\noindent Despite the several advances in the social sciences, we
have yet to discover an objective measuring stick for comparison,
a so called, True Comparison Theory, which can be an aid for arriving
at objective decisions. 
\end{doublespace}
\end{assumption}
\begin{doublespace}
\noindent The search for such a theory could be compared, to the medieval
alchemists’ obsession with turning everything into gold. For our present
purposes, the lack of such an objective measure means that the difference
in comparisons, as assessed by different participants, can effect
different decisions under the same set of circumstances. Hence, despite
all the uncertainty in the social sciences, the one thing we can be
almost certain about is the subjectivity in all decision making. This
lack of an objective measure for comparisons, makes people react at
varying degrees and at varying speeds, as they make their subjective
decisions. A decision gives rise to an action and subjectivity in
the comparison means differing decisions and hence unpredictable actions.
This inability to make consistent predictions in the social sciences
explains the growing trend towards comprehending better and deciphering
the decision process and the subsequent actions, by collecting more
information across the entire cycle of comparisons, decisions and
actions. 

\noindent Restricted to the particular sub-universe of economic and
financial theory, this translates to the lack of an objective measuring
stick of value, a so called, True Value Theory. This lack of an objective
measure of value, (hereafter, value will be synonymously referred
to as the price of a financial instrument), makes prices react at
differing degrees and at varying velocities to the pull of different
macro and micro factors. Another feature of the social sciences is
that the actions of participants affects the state of the system,
effecting a state transfer which perpetuates another merry-go-round
of comparisons, decisions and actions from the participants involved.
This means, more the participants, more the changes to the system,
more the actions and more the information that is generated to be
gathered. Hence perhaps, an unintended consequence of the recent developments
in technology has been to increase the complexity in our lives in
many ways.

\noindent The dynamic nature of the social sciences, where changes
can be observed and decisions can be taken by participants to influence
the system, means that along with better models and predictive technologies,
predictions need to be continuously revised; and yet unintended consequences
set in (Kashyap 2016); and as long as participants are free to observe
the results and modify their actions, this effect will persist. (Simon
1962) points out that any attempt to seek properties common to many
sorts of complex systems (physical, biological or social), would lead
to a theory of hierarchy since a large proportion of complex systems
observed in nature exhibit hierarchic structure; that a complex system
is composed of subsystems that, in turn, have their own subsystems,
and so on.

\noindent This might hold a clue to the miracle that our minds perform;
abstracting away from the dots that make up a picture, to fully visualizing
the image, that seems far removed from the pieces that give form and
meaning to it. To help us gain a better understanding of the relationships
between different elements of information, we might need a metric
built from smaller parts (Kashyap 2016 has a summary of these mathematical
tools), but gives optimal benefits when seen from a higher level.
Contrary to what conventional big picture conversations suggest, as
the spectator steps back and the distance from the picture increases,
the image becomes smaller yet clearer.
\end{doublespace}
\begin{doublespace}

\subsection{More Minds versus Some Machines}
\end{doublespace}

\begin{doublespace}
\noindent We currently lack a proper understanding of how, in some
instances, our brains (or minds; and right now it seems we don't know
the difference!) make the leap of learning from information to knowledge
to wisdom. With no disrespect to any adults, it would not be entirely
wrong to label children as better and faster learners than adults.
(Holt 2017) shows that in most situations our minds work best when
we use them in a certain way, and that young children tend to learn
better than grownups (and better than they themselves will when they
are older) because they use their minds in a special way, which is
a style of learning that fits their present condition. 
\end{doublespace}
\begin{problem}
\begin{doublespace}
\noindent Perhaps, the real challenge is to replicate the curiosity
and learning an infant displays (Reio Jr, Petrosko, Wiswell \& Thongsukmag
2006 discuss the measurement and conceptualization of curiosity; also
see: Loewenstein 1994; Loewy 1998; Berlyne 1954; 1966). Intellect
might be a byproduct of Inquisitiveness, demonstrating another instance
of an unintended yet welcome consequence. Collecting new pieces of
information is akin to curiosity in our agents from a software or
financial perspective. The emphasis here is on broad information sources,
that might not be seemingly related to finance, since many of the
best decision makers (inside and outside of the financial realm) have
diverse reading habits and eclectic tastes.
\end{doublespace}
\end{problem}
\begin{doublespace}
\noindent If ignorance is bliss, intrusion might just be the opposite
and bring misery. As the saying goes, Curiosity Terminated the Cat
and … (The movie Terminator should tell us about other unintended
consequences that might pop up in the AI adventure: Cameron \& Wisher
1991). This brings up the question of Art and Science in the practice
of asset management (and everything else in life?); which are more
related than we probably realize, ``Art is Science that we don't
know about; Science is Art restricted to a set of symbols governed
by a growing number of rules''. This frame of mind and approach to
seeking knowledge could be termed, Science without Borders, but combined
with the Arts. While the similarities between art and science, should
give us hope; we need to face the realities of the situation. Right
now, arguably, in most cases, we (including computers and intelligent
machines?) can barely make the jump from the information to the knowledge
stage; even with the use of cutting / (bleeding?) edge technology
and tools. This exemplifies three things: 
\end{doublespace}
\begin{enumerate}
\begin{doublespace}
\item We are still in the information age. As another route to establishing
this, consider this: Information is Hidden; Knowledge is Exchanged
or Bartered; Wisdom is Dispersed. Surely we are still in the Information
Age since a disproportionate amount of our actions are geared towards
accumulating unique data-sets for the sole benefits of the accumulators. 
\item Automating the movement to a higher level of learning, which is necessary
for dealing with certain doses of uncertainty, is still far away. 
\item Some of us missed the memo that the best of humanity are actually
robots in disguise, living amongst us.
\end{doublespace}
\end{enumerate}
\begin{doublespace}
\noindent Hence, it is not Manager versus Machine (Portfolio Manager
vs Computing Machine or MAN vs MAC, in short; for a discussion of
investment analysis, portfolio management and the building blocks
of modern computers, see: Reilly \& Brown 2002; Elton, Gruber, Brown
\& Goetzmann 2009; Bodie, Kane \& Marcus 2011; Perrier, Sipper \&
Zahnd 1996; Davis 2011; Amir, Ben-Ishay, Levner, Ittah, Abu-Horowitz
\& Bachelet 2014; Thompson, Gokler, Lloyd \& Shor 2016; End-notes
\ref{enu:Portfolio Manager}, \ref{enu:Universal Computing Machine},
\ref{enu:Computer}). Not even MAN and MAC against the MPC (Microsoft
Personal Computer: Freiberger \& Swaine 1999; Manes \& Andrews 1993;
Carlton \& Annotations-Kawasaki 1997; Wonglimpiyarat 2012; End-notes
\ref{enu: Mac or Macintosh}, \ref{enu:Personal Computer}, \ref{enu:MAC vs MPC})?
It is MAN, MAC and the MPC against increasing complexity! (Also in
scope are other computing platforms from the past, present and the
future: Williams 1997; Leuenberger \& Loss 2001; Ceruzzi 2003; Zhang,
Cheng \& Boutaba 2010; End-notes \ref{enu:History Computing}, \ref{enu:Computing Platform},
\ref{enu:Cloud Computing}, \ref{enu:Quantum Computing}). This increasing
complexity and information explosion is perhaps due to the increasing
number of complex actions perpetrated by the actors that comprise
the financial system. The human mind will be obsolete if machines
can fully manage assets and we would have bigger problems on our hands
than who is managing our money. We need, and will continue to need,
massive computing power to mostly separate the signal from the noise.
In this age of (Too Much) Information, it is imperative for Man and
Machine to work together to uncover nuggets of knowledge from buckets
of nonsense.
\end{doublespace}
\begin{doublespace}

\section{Buy Low / Sell High / Hold Now}
\end{doublespace}

\begin{doublespace}
\noindent To be clear, the requirement from the agent can be something
simple, like giving advice on a financial strategy. In this case,
the inputs can simply be the time series of numbers and the output
can be just a Buy, Sell or Hold indication, since all of finance through
time has involved only these three simple outcomes. The complications
are mostly to get to these three results, which the agent can conjure
up in its own way. But its interface with the external world, need
not be anything too involved.

\noindent The formal mathematical elements are discussed in (Kashyap
2017 has a road-map for creating intelligence). These quantitative
measures can be applied across aggregations of smaller elements that
can aid the AI agent by providing simple yet powerful metrics to compare
groups of entities and provide a relative valuation of which ones
are better candidates for acquisition and /or liquidation.
\end{doublespace}
\begin{doublespace}

\subsection{\label{subsec:Acing-the-Turing}Acing the Trader Turing Test }
\end{doublespace}

\begin{doublespace}
\noindent The Turing Test (TT) developed by Alan Turing (Turing 1950;
French 2000 chronicles the comments and controversy surrounding the
first fifty years of the TT; End-note \ref{enu:The-Turing-test}),
is a test of a machine's ability to exhibit intelligent behavior equivalent
to, or indistinguishable from, that of a human. Turing proposed that
a human evaluator would judge natural language conversations between
a human and a machine designed to generate human-like responses. The
evaluator would be aware that one of the two partners in conversation
is a machine, and all participants would be separated from one another.
The conversation would be limited to a text-only channel such as a
computer keyboard and screen so the result would not depend on the
machine's ability to render words as speech (Turing originally suggested
a teleprinter, one of the few text-only communication systems available
in 1950). If the evaluator cannot reliably tell the machine from the
human, the machine is said to have passed the test. The test does
not check the ability to give correct answers to questions, only how
closely answers resemble those a human would give.

\begin{figure}[H]
\includegraphics[width=17cm]{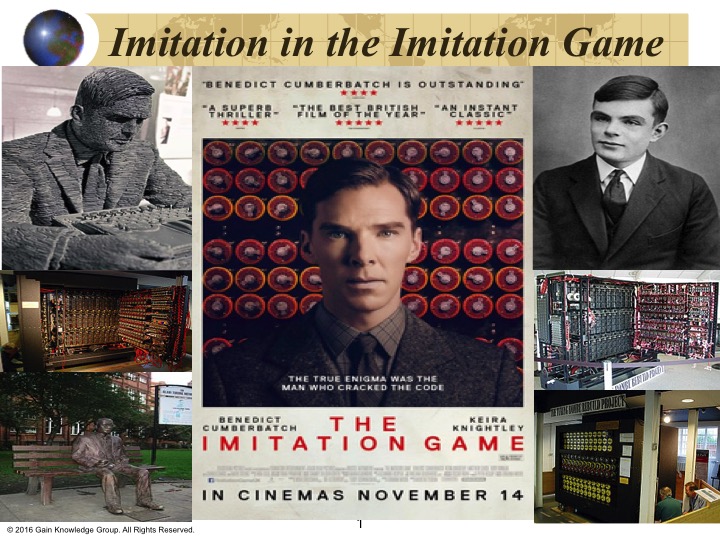}

\caption{\label{fig:The-Real-Enigma}The Real Enigma of the Imitation Game:
Which Alan Turing is the More Convincingly Brilliant Mathematician?}

\end{figure}

\noindent Let us now consider an example of imitation in the imitation
game (also known as, the Turing Test), which was also a recently released
movie about the role of Alan Turing in the second world war (You 2015).
The actor in the movie, Benedict Cumberbatch (Porter 2014), does a
marvelous job portraying the real Alan Turing (though this is a subjective
evaluation, if someone disagrees, termed a disbeliever, then it would
be fair to state that they now have the responsibility of doing a
better role play). The manner in which Benedict Cumberbatch plays
the main character in the movie, Imitation Game, leads us to state
the Real Enigma of the Imitation Game as: Which Alan Turing is the
More Convincingly Brilliant Mathematician? (Figure \ref{fig:The-Real-Enigma};
End-note \ref{enu:Most-people-when-Cumberbatch}). This question merely
inquires as to whether, Alan Turing or Benedict Cumberbatch, would
pass a stage test for actors who had to convince the audience they
were mathematicians. To go into length on how Benedict Cumberbatch
(or any disbeliever, forced to turn into a better actor) accomplished
this, would require another paper or a few books of their own (Hagen
1991); the short answer would be that, an actor believes that he can
play the part he is chosen to play, which is what an agent chosen
to display intelligence, must first be made to believe. This is about
not about dishonesty or deception, it is about belief and confidence.
True confidence comes when we admit we don't know something and we
are willing to try as discussed in section \ref{subsec:I-Don't-Know,}.

\noindent 

\noindent The Turing test for a trader would involve less interaction
than the regular Turing test, but it would require constantly assessing
some quantitative measures. We could use a combination of commonly
used portfolio management measures such as the Sharpe ratio, max draw-down
or maximum loss amount, Value at Risk and so on. (Cartea \& Jaimungal
2015) propose risk metrics to assess the performance of HFT strategies
that seek to maximize profits from making the realized spread where
the holding period is extremely short (fractions of a second, seconds,
or at most minutes; (De Prado \& Peijan 2004) measure the loss potential
of hedge funds; (Fung \& Hsieh 2004) is a model of hedge fund returns
based on dynamic risk-factor coefficients; (Evans 2004) discuss investor
attitudes toward risk, expectations of future portfolio returns, and
behaviors in the use of risk metrics; (for more discussions on financial
performance measurement and risk, see: Bacidore, Boquist, Milbourn
\& Thakor 1997; Ferguson \& Leistikow 1998; Donefer 2010; Giot \&
Laurent 2003; Treleaven, Galas \& Lalchand 2013).

\noindent An often omitted criteria that needs to be considered when
administering the Turing test is the ability, or, the level of skill,
of the person conducting the test. Surely, different individuals are
satisfied with different levels of impersonation. In a trading context,
a perfect role model for our agent to mimic, would perhaps be Michael
Douglas from the movie Wall Street (Formaini 2001; Paddock, Terranova
\& Giles 2001). To be precise, the automation, would certainly fail
to capture the many nuances to being Michael Douglas. But all it would
need to do is exhibit his confidence when making financial decisions
(this could be captured by how accurate the confidence intervals from
probability theory given by the agent are: Rao 1973; Cramér 2016).

\noindent When we see any drama, play or movie that depicts the life
of any real trader, (while reminding ourselves that movies might not
be real, but real life can become movies: The Wolf of Wall Street
is a great example, Belfort 2011; End-note \ref{enu:The-Wolf-Wall-Street});
different people are satisfied with different levels of acting ability.
We all know, that the person playing the role in the theatrical version
is not the same, as the person, that is being enacted. But in many
cases, (perhaps, in most cases, when it is well produced), we leave
feeling satisfied with the result of the replication. The lesson for
us here is this: how far does the test administrator need to go, to
believe that the computer program perfectly duplicates a human trader
or is able to surpass the quantitative benchmarks that have been set.

\noindent Attempts to chronicle what makes a good trader usually result
in many interesting and varied findings. A consensus would perhaps
be around the fact that better traders are better risk managers and
focus their efforts on more actively looking after the downside since
the upside usually takes care of itself. This results in reduced variance
of the Profit \& Loss (P\&L) of the portfolio. Another viewpoint is
that good traders can make bad trades and bad traders can make good
trades, but over many iterations, the good traders end up making a
greater number of good trades and hence we provide some distinctions
between good and bad traders (trades) in section \ref{subsec:Characteristics-of-a}
(Kashyap 2014).
\end{doublespace}
\begin{doublespace}

\subsection{\label{subsec:Characteristics-of-a}Good Traders, Bad Trades ...
Ugly Repercussions}
\end{doublespace}
\begin{enumerate}
\begin{doublespace}
\item The factors that dictate a good trader or a bad trader depend on the
Time Horizon and the Investment Objective. The time horizon can be
classified into short term, medium term and long term. The investment
objective can be conservative or aggressive. While there are no strict
boundaries between these categories, such a classification helps us
with the analysis and better identification of traders.
\item Any trader that fulfills the investment objective and time horizon
for which his trades are made is a good trader. Otherwise, he / she
(it, in the case of artificial agents?) is a bad trader. 
\item On the face of it, we can view good traders as the profitable ones
and bad traders as ones that lose money. But where possible, if we
try and distinguish between proximate causes and ultimate reasons,
it becomes apparent that good traders can lose money and bad traders
can end up making money. 
\item As discussed in sections \ref{sec:The-Circle-of}, \ref{subsec:Automation-for-What},
due to the nature of uncertainty in the social sciences: the noise
around the expected performance of any security; our ignorance of
the true equilibrium; the behavior of other participants; risk constraints
like liquidity, concentration, unfavorable Geo-political events, etc.
implies we would have deviations from our intended results. The larger
the deviation from the intended results, the worse our trader is.
\item What the above implies is that, bad traders show the deficiencies
in planning (estimation process) and how we have not been able to
take into account factors that can lead our results astray. It is
true that due to the extreme complexity of the financial markets,
the unexpected ends up happening and we can never take into account
everything. We just need to make sure that the unexpected, even if
it does happen, is contained in the harm it can cause. The good thing
about bad traders (trades) is the extremely valuable lessons they
hold for us, which takes us through the loop or trials, errors and
improvements. 
\item We then need to consider, how a good trader can lose money. When we
make a trade, if we know the extent to which we can lose, when this
loss can occur and that situation ends up happening, our planning
did reveal the possibility and extent of the loss, hence it is a good
trade.
\item The bottom line is that, good traders (trades), or bad traders (trades),
are the result of our ability to come up with possible scenarios and
how likely we think they will happen.
\end{doublespace}
\end{enumerate}
\begin{doublespace}

\subsection{\label{subsec:Mexican-Chihuahua-solving}Mexican Chihuahua Trading
Korean Bonds under a Mush-Room}
\end{doublespace}

\begin{doublespace}
\noindent John Searle’s thought experiment (Searle 1980; Preston \&
Bishop 2002 has a collection of essays on this crucial challenge;
End-note \ref{enu:Chinese room thought experiment}) begins with this
hypothetical premise: suppose that artificial intelligence research
has succeeded in constructing a computer that behaves as if it understands
Chinese. It takes Chinese characters as input and, by following the
instructions of a computer program, produces other Chinese characters,
which it presents as output. Suppose, says Searle, that this computer
performs its task so convincingly that it comfortably passes the Turing
test: it convinces a human Chinese speaker that the program is itself
a live Chinese speaker. To all of the questions that the person asks,
it makes appropriate responses, such that any Chinese speaker would
be convinced that they are talking to another Chinese-speaking human
being. This was originally phrased as: ``Searle supposes that he
is in a closed room and has a book with an English version of the
computer program, along with sufficient paper, pencils, erasers, and
filing cabinets. Searle could receive Chinese characters through a
slot in the door, process them according to the program's instructions,
and produce Chinese characters as output''.

\noindent The question Searle wants to answer is this: does the machine
literally \textquotedbl understand\textquotedbl{} Chinese? Or is
it merely simulating the ability to understand Chinese? Searle calls
the first position \textquotedbl strong AI\textquotedbl{} and the
latter \textquotedbl weak AI\textquotedbl . Searle writes that \textquotedbl according
to Strong AI, the correct simulation really is a mind. According to
Weak AI, the correct simulation is a model of the mind.\textquotedbl{}
He also writes: \textquotedbl On the Strong AI view, the appropriately
programmed computer does not just simulate having a mind; it literally
has a mind.\textquotedbl{}

\noindent Searle was in fact against the notion of strong AI, which
is that human minds are, in essence, computer programs. That is an
appropriately programmed computer with the right inputs and outputs,
would thereby have a mind in exactly the same sense human beings have
minds. All mental activity, is simply the carrying out of some well-defined
sequence of operations, frequently referred to, as an algorithm. 

\noindent Searle’s example has had a profound impact on the discussions
related to AI for the last many years. However, as a counter argument
let us consider, instead of an American (John Searle) juggling with
Chinese characters he has no clue about in a closed room, using instructions
in English, a language he understands; what if it was a Mexican Chihuahua
making decisions on a portfolio of Korean bonds, (the inputs it receives
are numbers or time series of prices, other variables and perhaps
other information related to the financial markets in Korean characters),
sitting under a giant Mush-Room (perhaps, having devoured the mushroom
and hence being influenced by it in ways, that we do not quite yet
comprehend, but for the purposes of this test, the effects are only
beneficial. For the hallucinogenic effects of mushrooms see: Schwartz
\& Smith 1988), and giving out the right answers back in the form
of Korean characters (the output it provides could simply be buy,
sell or hold characters in Korean or a full summary or justification
of the investment decision as well) and has a remarkably good track
record in terms of risk and returns, but only barks in response to
everything else. 

\noindent Does it matter, whether the Chihuahua is only using certain
training it has been given, to use rules to arrange Korean characters
as excellent investment advice, or, whether it is the Mushroom causing
the miracle or something else? For all practical purposes, the Chihuahua
is a great portfolio manager, that simply does not speak the same
language as we do. We do not understand its barking nor does it understand
the voice tones we produce, or maybe it pretends that it does not
understand (it can be argued, though we won’t continue this line of
reasoning, that we understand less of what dogs say, than what dogs
understand of what we say; who is more intelligent then?). For simplicity
and for rhetorical reasons, let us just say that the effects of the
mushroom, last for as long as the Chihuahua is alive, or, until we
are still interested in asking it questions about portfolio management
using Korean Characters? 

\noindent To substantiate this counter viewpoint, does it really matter
if we are simply using rules to trade or if we are actually understanding
how the investment choice was arrived at, if we completely believe
that we understand the solution and merely use rules to arrive at
the solution? This is not about being dishonest, or, passing lie detector
tests; since, if we believe we can generate profits and if we are
able to consistently generate profits, it does not matter, how we
got the profits, since we should now be deemed intelligent enough,
as we have come up with the profits.
\end{doublespace}
\begin{doublespace}

\subsection{\label{subsec:I-Don't-Know,}I Don't Know, A Great Answer}
\end{doublespace}

\begin{doublespace}
\noindent (Taleb 2007) in his landmark book, the Black Swan, talks
about the unread books in the personal library of legendary Italian
writer, Umberto Eco, and how over time, this unread collection gets
larger. Hence, it would not be incorrect, to say that, there is more
that, we don’t know, than, what we know; the more we know, the more,
there will be to know. But that should not stop us and the agent,
from trying to seek the answers, or, even from making a guess, as
a starting point.

\noindent Hence, an answer admitting, ``I Don't Know'' is a great
answer in most situations. When we design any system or model, especially
in AI, questions and answers are important, since that is the primary
way to assess the presence of intelligence. But what becomes more
important are our definitions and assumptions. To supplement our definition
of intelligence, we provide the following cardinal assumption,
\end{doublespace}
\begin{assumption}
\begin{doublespace}
\noindent \label{assu:The-knowledge-that}The knowledge that has been
accumulated over time is lesser than the knowledge that is yet to
gathered. With this assumption, an answer of ``I Don't Know'' becomes,
not just a correct answer, but it is an invitation to the person asking
the question to teach the agent how to answer the question.
\end{doublespace}
\end{assumption}
\begin{doublespace}
\noindent So the agent is always learning, and the reason is, simply,
due to what we discussed before: we don’t know most things and hence,
the learning, usually never stops. If the person asking the question
is not satisfied with the answer, he or she, now has a responsibility
to teach the agent, to improve upon the answer produced. A failure
to create intelligence in any agent is a failure on the part of the
teacher in finding a teaching methodology, appropriate for the agent.
This also implies that: 
\end{doublespace}
\begin{criterion}
\begin{doublespace}
\noindent \label{cri:Creating-intelligence-is}Creating intelligence
is not only about writing software code, it is about having the best
teachers that humanity has produced, being available to teach the
later generations, be it human or machines. In a trading context,
to create intelligent trading machines, we need, not just the best
programmers, but we also need the best traders, portfolio strategists
and risk managers, to teach the agents, how to come up with investment
ideas, implement them and manage the resulting risk.
\end{doublespace}
\end{criterion}
\begin{doublespace}
\noindent We now consider the fundamental question of whether we need
complicated models or merely stronger beliefs. We state this as our
essential doctrine.
\end{doublespace}
\begin{criterion}
\begin{doublespace}
\noindent \label{cri:The-agent-has-Self-Confidence}The intelligent
agent has to believe that it has the ability to learn, and the confidence
to request lessons regarding answers, that it is unable to generate
satisfactorily.
\end{doublespace}
\end{criterion}
\begin{doublespace}
\noindent Combining confidence with the great answer, which follows
from our assumption \ref{assu:The-knowledge-that}, we get a better
answer, which is ``let me try''. 

\noindent When an agent is not learning, it should ideally be teaching
(other agents or anyone else). Perhaps because, teaching and learning
are highly interconnected and the best way to learn is to teach. A
realization that the roles of students and teachers, are constantly
getting interchanged, originates from a belief that, everyone has
something to teach, to everyone else. When we are teaching, we are
also learning from someone else, when we are learning, we are really
teaching ourselves. To be clear, although, most of us probably know
this, learning does not just represent, reading textbooks, or, doing
assignments, though, these are important components of learning. Learning
can happen, when we are doing anything, that we enjoy doing. This
can be built into the reward system of the agent, so that it accumulates
points for aspects that it likes. Different agents could be made to
like different things, so that we build a random enjoyment component
that learns from different activities.

\noindent Efforts at learning and teaching, usually end up confronting
two monsters: Confusion and Frustration, both of which, though, scary
and ugly to begin with, can be powerful motivators, as long as, we
don’t let them bother us. Confusion is the beginning of Understanding.
Necessity, is the mother of all creation / innovation / invention,
but the often forgotten father, is Frustration. What we learn from
the story of, Beauty and the Beast, is that, we need to love the beasts
to find beauty. Hence, if we start to love these monsters (Confusion
and Frustration), we can unlock their awesomeness and find truly stunning
solutions.

\noindent Hence, our agent has to remain confident and ask questions,
when it does not have an answer. This can also be stated as, 
\end{doublespace}
\begin{sol}
\begin{doublespace}
\noindent \textit{\label{cond:Life-for-an}Life for an intelligent
agent is all about having confidence and the right teachers and /or
students.}
\end{doublespace}
\end{sol}
\begin{doublespace}

\subsection{\label{subsec:Merry-Go-Round-of-Decisions,}Merry-Go-Round of Trials,
Errors and Revisions}
\end{doublespace}

\begin{doublespace}
\noindent Usually, on our first attempt to answer any question, we
may not get the correct, or, the best answer. This is where, the trial
and error part, kicks in. But once, we start somewhere, we learn from
our mistakes and improve upon our explanations. In this Question \&
Answer context, we define any question as a good question and a good
answer as something that we only think of later or something we find
after a few iterations of trial and error.

\noindent (Young 2009) is about trial and learning in a social or
economic game theory setting (Gibbons 1992). A person learns by trial
and error if he occasionally tries out new strategies, rejecting choices
that are, erroneous, in the sense that they do not lead to higher
payoffs. In an economic game, however, strategies can become erroneous
due to a change of behavior by someone else, triggering a search for
new and better strategies. In economics, it is insightful to establish
conditions under which the Nash equilibrium property (Nash 1950) can
be established. But in real life, equilibrium is a dynamic, constantly
changing state due to the subjectivity in all decision making and
the differing perceptions of the individuals involved and hence the
trial and error never ceases. 

\noindent In all efforts at creating intelligence, we make an unstated
assumption that human beings are capable of intelligence. But, we
are not born intelligent (maybe we are, but, perhaps, we just don't
know). It takes years of nurturing and tutoring for us to become intelligent
and we display different abilities and aptitude for different things,
or the intelligence of different individual could be in different
skills. How could we then have expectations that something, that we
deem not to have the capacity for intelligence, has to become intelligent
in a relatively short span of time. This holds a strong message for
us that, to create intelligence artificially, might require years
of training for an agent.

\noindent In a typical classroom, some kids end up with more conventional
forms of intelligence in comparison to others and as assessed by our
benchmark or measure, due to creating more connections and retaining
the relevant bits of information they receive. Hence, we would expect
a similar sort of situation when trying to create AI, we need to start
with a group of agents, with different parameters and let them wander
around and see what innate abilities they pick up. Accordingly, we
need to further those skills that were naturally acquired. The circle
of trial, error and corrections needs to be happening constantly.
\end{doublespace}
\begin{doublespace}

\section{\label{sec:A-Journey-to}A Journey to the Land of Unintended Consequences}
\end{doublespace}

\begin{doublespace}
\noindent We have discussed the intuition for why we need the best
teachers and not just the best computing science designers in creating
automated trading systems. An unintended consequence of establishing
curiosity and confidence in an agent, expected to become intelligent,
might well be intelligence. We have considered why, even though we
wish to create intelligence and make the agent pass tests of intelligence,
the gift of intelligence might be something from the realm of the
unintentional. The mathematical tools and formal elements of what
such an endeavor might require, which includes models of diffusion,
distance measures and dimension reduction, among other things are
discussed in detail in (Kashyap 2017 has a how to guide for creating
intelligence) and briefly mentioned in Appendix \ref{sec:Appendix-A:-From}. 
\end{doublespace}
\begin{lyxalgorithm}
\begin{doublespace}
\noindent We summarize the formal model, (mathematics being the language
of ceremoniously concise precision) using words in the dictionary
as below. Each following point could also be implemented as separate
computer programs with different parameters that govern the behavior
of the agent. Testing could be done independently and varying parameters
can be used for combining the elements below as well.
\end{doublespace}
\end{lyxalgorithm}
\begin{rem}
\begin{doublespace}
\noindent \textbf{\textit{\textcolor{black}{We could consider each
point below as a component that can be tested as a separate scientific
hypothesis (relevant references are given in the corresponding sections);
but surely, greater the consonance between the ingredients that encapsulate
the below concepts, better the overall outcome.}}}
\end{doublespace}
\end{rem}
\begin{enumerate}
\begin{doublespace}
\item Information collection is approximated using the Bass model of diffusion
(section \ref{subsec:Bass-Model-of}), which is used extensively in
marketing to study the adoption of new products. Collecting new pieces
of information is how we mimic curiosity in our agents.
\item \label{enu:Information-accumulated-is}Information accumulated is
stored and periodically the elements gathered are compared to establish
how strong the connections between the elements are. When solving
problems, we want to rely on bits of information, that are not too
closely related, but that are also not significantly unrelated. This
requires a trial and error approach, with different parameters across
multiple agents, as discussed in section \ref{subsec:Merry-Go-Round-of-Decisions,}.
\item To perform the comparison in step \ref{enu:Information-accumulated-is},
we will need to use the Bhattacharyya Distance and Johnson-Lindenstrauss
Lemma (sections \ref{subsec:Bhattacharyya-Distance}, \ref{subsec:Dimension-Reduction}).
\end{doublespace}
\end{enumerate}
\begin{doublespace}
\noindent A glimpse, of what a journey towards the land of unintended
consequences holds, can be seen, by reminding ourselves that all profits
(P\&L) generation, is but an unintended consequence. Although, to
be precise, traders, do want to intentionally make profitable trades,
but the exact new trades that end up becoming profitable are unintentional;
they stumble upon it, as they wander around the financial landscape,
putting on new positions and managing their risk.

\noindent Success is, a very relative term. In the extreme case, which
is relevant in finance, sometimes, one person’s success (profit) could
be someone else’s failure (loss). That being said, to triumph in creating
automated strategies and almost everything else, it is important to
know where we are, and start the journey towards, where we want to
be. This happens by putting on trades, under the guidance of a better
trader, who will tutor the agents on why certain suggestions it has
made need to be revised. Such lessons need to be reflected in the
knowledge store of the agent.

\noindent An unintended consequence of taking the first step on a
journey, means that the percentage of the distance left to be traveled
reduces from infinity to a finite number. So once we start the trip,
it becomes manageable immediately. The subjectivity in how we compare
things, means that the benchmark for automated trading will be constantly
changing, which means, we need our agents to keep on learning, just
as we need to do the same, as well.
\end{doublespace}
\begin{doublespace}

\section{\label{sec:Appendix-A:-From}Appendix A: From Words to Symbols, A
Curious and Confident Trader Model}
\end{doublespace}
\begin{doublespace}

\subsection{Notation and Terminology for Key Results}
\end{doublespace}
\begin{itemize}
\begin{doublespace}
\item $D_{BC}\left(p_{i},p_{i}^{\prime}\right)$, the Bhattacharyya Distance
between two multinomial populations each consisting of $k$ categories
classes with associated probabilities $p_{1},p_{2},...,p_{k}$ and
$p_{1}^{\prime},p_{2}^{\prime},...,p_{k}^{\prime}$ respectively.
\item $\rho\left(p_{i},p_{i}^{\prime}\right)$, the Bhattacharyya Coefficient.
\item $D_{BC-N}(p,q)$ is the Bhattacharyya distance between $p$ and $q$
normal distributions or classes.
\item $D_{BC-MN}\left(p_{1},p_{2}\right)$ is the Bhattacharyya distance
between two multivariate normal distributions, $\boldsymbol{p_{1}},\boldsymbol{p_{2}}$
where $\boldsymbol{p_{i}}\sim\mathcal{N}(\boldsymbol{\mu}_{i},\,\boldsymbol{\Sigma}_{i})$.
\item $D_{BC-TN}(p,q)$ is the Bhattacharyya distance between $p$ and $q$
truncated normal distributions or classes.
\item $D_{BC-TMN}\left(p_{1},p_{2}\right)$ is the Bhattacharyya distance
between two truncated multivariate normal distributions, $\boldsymbol{p_{1}},\boldsymbol{p_{2}}$
where $\boldsymbol{p_{i}}\sim\mathcal{N}(\boldsymbol{\mu}_{i},\,\boldsymbol{\Sigma}_{i},\,\boldsymbol{a}_{i},\,\boldsymbol{b}_{i})$.
\end{doublespace}
\end{itemize}
\begin{doublespace}

\subsection{\label{subsec:Bass-Model-of}Bass Model of Diffusion for Information
Accumulation}
\end{doublespace}

\begin{doublespace}
\noindent Collecting new pieces of information is the behavioral parallel
we draw to creating curiosity in our agents. We model collection of
information, using the Bass Model of Diffusion, which is used extensively
in the marketing field to model the adoption of new products by consumers.
One of the simplest forms of the Bass model and also the original
one from the pioneer (Bass 1969; End-note \ref{enu:Bass-Model-Diffusion})
can be written as,
\[
\frac{f\left(t\right)}{1-F\left(t\right)}=p+qF\left(t\right)
\]
\[
F\left(t\right)=\int_{0}^{t}f\left(u\right)du
\]
Here,

\noindent $f\left(t\right)$, is the change of the installed base
fraction or the likelihood of purchase at time $t$.

\noindent $F\left(t\right)$, is the installed base fraction.

\noindent $p$, is the coefficient of innovation.

\noindent $q$, is the coefficient of imitation.

\noindent Sales $S\left(t\right)$ is the rate of change of installed
base (i.e. adoption), that is, $f\left(t\right)$ multiplied by the
ultimate market potential $m$. This is given by,
\[
S\left(t\right)=mf\left(t\right)
\]
\[
S\left(t\right)=m\frac{\left(p+q\right)^{2}}{p}\frac{e^{-\left(p+q\right)t}}{\left(1+\frac{q}{p}e^{-\left(p+q\right)t}\right)^{2}}
\]

\noindent We view a new product being adopted as being equivalent
to the agent collecting the adopter, which is the new piece of information
that got collected. This could represent the time series of prices,
or the balance sheet strength of a company, or any other element of
information that could be useful towards investment decisions. Though
the information need not just be restricted to finance, but could
be anything related (or seemingly unrelated) and potentially useful
to provide a vivid summary of the world we live in and eventually
contribute to an uptick in the overall intelligence.

\noindent As alternative models, we could use models used in economics
for the spread of rumors (Banerjee 1993 has a discussion of information
transmission processes, which for our purposes are similar to information
collection processes) and herd behavior (Christie \& Huang 1995; Chiang
\& Zheng 2010; Chen 2013; Muñoz Torrecillas, Yalamova \& McKelvey
2016 are about herding behavior in financial markets). (Banerjee 1992)
is a sequential decision model in which each decision maker looks
at the decisions made by previous decision makers in taking her own
decision, showing that the decision rules that are chosen by optimizing
individuals will be characterized by herd behavior; i.e., people will
be doing what others are doing rather than using their information.
\end{doublespace}
\begin{doublespace}

\subsection{\label{subsec:Bhattacharyya-Distance}Bhattacharyya Distance for
Information Comparison}
\end{doublespace}

\begin{doublespace}
\noindent We use the Bhattacharyya distance (Bhattacharyya 1943; 1946)
as a measure of similarity or dissimilarity between the probability
distributions of the two entities we are looking to compare. These
entities could be two information sources, two securities, groups
of securities, markets or any statistical populations that we are
interested in studying. The Bhattacharyya distance is defined as the
negative logarithm of the Bhattacharyya coefficient. 
\[
D_{BC}\left(p_{i},p_{i}^{\prime}\right)=-\ln\left[\rho\left(p_{i},p_{i}^{\prime}\right)\right]
\]
The Bhattacharyya coefficient is calculated as shown below for discrete
and continuous probability distributions. 
\[
\rho\left(p_{i},p_{i}^{\prime}\right)=\sum_{i}^{k}\sqrt{p_{i}p_{i}^{\prime}}
\]
\[
\rho\left(p_{i},p_{i}^{\prime}\right)=\int\sqrt{p_{i}\left(x\right)p_{i}^{\prime}\left(x\right)}dx
\]

\noindent Bhattacharyya’s original interpretation of the measure was
geometric (Derpanis 2008). He considered two multinomial populations
each consisting of $k$ categories classes with associated probabilities
$p_{1},p_{2},...,p_{k}$ and $p_{1}^{\prime},p_{2}^{\prime},...,p_{k}^{\prime}$
respectively. Then, as $\sum_{i}^{k}p_{i}=1$ and $\sum_{i}^{k}p_{i}^{\prime}=1$,
he noted that $(\sqrt{p_{1}},...,\sqrt{p_{k}})$ and $(\sqrt{p_{1}^{\prime}},...,\sqrt{p_{k}^{\prime}})$
could be considered as the direction cosines of two vectors in $k-$dimensional
space referred to a system of orthogonal co-ordinate axes. As a measure
of divergence between the two populations Bhattacharyya used the square
of the angle between the two position vectors. If $\theta$ is the
angle between the vectors then: 
\[
\rho\left(p_{i},p_{i}^{\prime}\right)=cos\theta=\sum_{i}^{k}\sqrt{p_{i}p_{i}^{\prime}}
\]
Thus if the two populations are identical: $cos\theta=1$ corresponding
to $\theta=0$, hence we see the intuitive motivation behind the definition
as the vectors are co-linear. Bhattacharyya further showed that by
passing to the limiting case a measure of divergence could be obtained
between two populations defined in any way given that the two populations
have the same number of variates. The value of coefficient then lies
between $0$ and $1$. 
\[
0\leq\rho\left(p_{i},p_{i}^{\prime}\right)\leq=1
\]
\[
0\leq D_{BC}\left(p_{i},p_{i}^{\prime}\right)\leq\infty
\]
We get the following formulae (Lee and Bretschneider 2012) for the
Bhattacharyya distance when applied to the case of two uni-variate
normal distributions. 
\[
D_{BC-N}(p,q)=\frac{1}{4}\ln\left(\frac{1}{4}\left(\frac{\sigma_{p}^{2}}{\sigma_{q}^{2}}+\frac{\sigma_{q}^{2}}{\sigma_{p}^{2}}+2\right)\right)+\frac{1}{4}\left(\frac{(\mu_{p}-\mu_{q})^{2}}{\sigma_{p}^{2}+\sigma_{q}^{2}}\right)
\]

\noindent $\sigma_{p}$ is the variance of the $p-$th distribution, 

\noindent $\mu_{p}$ is the mean of the $p-$th distribution, and 

\noindent $p,q$ are two different distributions.

\noindent The original paper on the Bhattacharyya distance (Bhattacharyya
1943) mentions a natural extension to the case of more than two populations.
For an $M$ population system, each with $k$ random variates, the
definition of the coefficient becomes, 
\[
\rho\left(p_{1},p_{2},...,p_{M}\right)=\int\cdots\int\left[p_{1}\left(x\right)p_{2}\left(x\right)...p_{M}\left(x\right)\right]^{\frac{1}{M}}dx_{1}\cdots dx_{k}
\]

\noindent For two multivariate normal distributions, $\boldsymbol{p_{1}},\boldsymbol{p_{2}}$
where $\boldsymbol{p_{i}}\sim\mathcal{N}(\boldsymbol{\mu}_{i},\,\boldsymbol{\Sigma}_{i}),$
\[
D_{BC-MN}\left(p_{1},p_{2}\right)=\frac{1}{8}(\boldsymbol{\mu}_{1}-\boldsymbol{\mu}_{2})^{T}\boldsymbol{\Sigma}^{-1}(\boldsymbol{\mu}_{1}-\boldsymbol{\mu}_{2})+\frac{1}{2}\ln\,\left(\frac{\det\boldsymbol{\Sigma}}{\sqrt{\det\boldsymbol{\Sigma}_{1}\,\det\boldsymbol{\Sigma}_{2}}}\right),
\]

\noindent $\boldsymbol{\mu}_{i}$ and $\boldsymbol{\Sigma}_{i}$ are
the means and covariances of the distributions, and $\boldsymbol{\Sigma}=\frac{\boldsymbol{\Sigma}_{1}+\boldsymbol{\Sigma}_{2}}{2}$.
We need to keep in mind that a discrete sample could be stored in
matrices of the form $A$ and $B$, where, $n$ is the number of observations
and $m$ denotes the number of variables captured by the two matrices.
\[
\boldsymbol{A_{m\times n}}\sim\mathcal{N}\left(\boldsymbol{\mu_{1}},\boldsymbol{\varSigma_{1}}\right)
\]
\[
\boldsymbol{B_{m\times n}}\sim\mathcal{N}\left(\boldsymbol{\mu_{2}},\boldsymbol{\varSigma_{2}}\right)
\]

\end{doublespace}
\begin{doublespace}

\subsection{\label{subsec:Dimension-Reduction}Dimension Reduction before Information
Comparison}
\end{doublespace}

\begin{doublespace}
\noindent A key requirement to apply the Bhattacharyya distance in
practice is to have data-sets with the same number of dimensions.
(Fodor 2002; Burges 2009; Sorzano, Vargas and Montano 2014) are comprehensive
collections of methodologies aimed at reducing the dimensions of a
data-set using Principal Component Analysis or Singular Value Decomposition
and related techniques. (Johnson and Lindenstrauss 1984) proved a
fundamental result (JL Lemma) that says that any $n$ point subset
of Euclidean space can be embedded in $k=O(log\frac{n}{\epsilon^{2}})$
dimensions without distorting the distances between any pair of points
by more than a factor of $\left(1\pm\epsilon\right)$, for any $0<\epsilon<1$.
Whereas principal component analysis is only useful when the original
data points are inherently low dimensional, the JL Lemma requires
absolutely no assumption on the original data. Also, note that the
final data points have no dependence on $d$, the dimensions of the
original data which could live in an arbitrarily high dimension. We
use the version of the bounds for the dimensions of the transformed
subspace given in (Frankl and Maehara 1988; 1990; Dasgupta and Gupta
1999).
\end{doublespace}
\begin{lem}
\begin{doublespace}
\noindent \label{Prop:Johnson and Lindenstrauss --- Dasgupta and Gupta}For
any $0<\epsilon<1$ and any integer $n$, let $k$ be a positive integer
such that 
\[
k\geq4\left(\frac{\epsilon^{2}}{2}-\frac{\epsilon^{3}}{3}\right)^{-1}\ln n
\]
Then for any set $V$ of $n$ points in $\boldsymbol{R}^{d}$, there
is a map $f:\boldsymbol{R}^{d}\rightarrow\boldsymbol{R}^{k}$ such
that for all $u,v\in V$, 
\[
\left(1-\epsilon\right)\Vert u-v\Vert^{2}\leq\Vert f\left(u\right)-f\left(v\right)\Vert^{2}\leq\left(1+\epsilon\right)\Vert u-v\Vert^{2}
\]
Furthermore, this map can be found in randomized polynomial time and
one such map is $f=\frac{1}{\sqrt{k}}Ax$ where, $x\in\boldsymbol{R}^{d}$
and $A$ is a $k\times d$ matrix in which each entry is sampled i.i.d
from a Gaussian $N\left(0,1\right)$ distribution.
\end{doublespace}
\end{lem}
\begin{doublespace}
\noindent (Kashyap 2016) provides expressions for the density functions
after dimension transformation when considering log normal distributions,
truncated normal and truncated multivariate normal distributions (Norstad
1999). These results are applicable in the context of many variables
observed in real life such as stock prices, trading volumes, sales
or inventory levels, and volatilities, which do not take on negative
values. We also require the expression for the dimension transformed
normal distribution, since it is a better candidate to model returns.
which could take on negative values. A relationship between covariance
and distance measures is also derived. We point out that these mathematical
concepts have many uses outside the domain of finance.
\end{doublespace}
\begin{doublespace}

\section{\label{sec:Sleeping-Aids}End-notes}
\end{doublespace}
\begin{enumerate}
\begin{doublespace}
\item Numerous seminar participants suggested ways to improve the manuscript.
The views and opinions expressed in this article, along with any mistakes,
are mine alone and do not necessarily reflect the official policy
or position of either of my affiliations or any other agency.
\item \label{enu:The-Red-Queen's}The Red Queen's race is an incident that
appears in Lewis Carroll's Through the Looking-Glass and involves
the Red Queen, a representation of a Queen in chess, and Alice constantly
running but remaining in the same spot.
\end{doublespace}

\begin{doublespace}
\noindent \textquotedbl Well, in our country,\textquotedbl{} said
Alice, still panting a little, \textquotedbl you'd generally get
to somewhere else, if you run very fast for a long time, as we've
been doing.\textquotedbl{}

\noindent \textquotedbl A slow sort of country!\textquotedbl{} said
the Queen. \textquotedbl Now, here, you see, it takes all the running
you can do, to keep in the same place. If you want to get somewhere
else, you must run at least twice as fast as that!\textquotedbl{}

\noindent \href{https://en.wikipedia.org/wiki/Red_Queen\%27s_race}{The Red Queen's Race, Wikipedia Link}

\noindent This quote is commonly attributed as being from Alice in
Wonderland as: “My dear, here we must run as fast as we can, just
to stay in place. And if you wish to go anywhere you must run twice
as fast as that.”
\end{doublespace}
\begin{doublespace}
\item \label{enu:Most-people-when-Cumberbatch} Most people when posed the
question: ``Is Benedict Cumberbatch a mathematician of extraordinary
ability?'', would answer in the negative. This answer comes about,
without most of us having met him, or, knowing whether he has been
studying mathematics secretly for years but not having obtained any
formal degree in the field; again highlighting, how we jump to conclusions.
Unlikely as it seems, it is still probabilistically possible that
he might be an exemplary mathematician. It is worth pointing out that
most of us have this belief about Mr. Cumberbatch, without even knowing
what his educational background is; though in this case, there are
no surprises since, a quick search on Wikipedia or Google \href{https://en.wikipedia.org/wiki/Benedict_Cumberbatch}{Benedict Cumberbatch, WIkipedia Link}
will reveal that his many years of formal training have been in acting.
\item \label{enu:HFT}\href{https://en.wikipedia.org/wiki/High-frequency_trading}{HFT, High Frequency Trading, Wikipedia Link}
In financial markets, high-frequency trading (HFT) is a type of algorithmic
trading characterized by high speeds, high turnover rates, and high
order-to-trade ratios that leverages high-frequency financial data
and electronic trading tools (Aldridge 2013). While there is no single
definition of HFT, among its key attributes are highly sophisticated
algorithms, co-location, and very short-term investment horizons.
HFT can be viewed as a primary form of algorithmic trading in finance
(End-note \ref{enu:Algorithmic Trading}). Specifically, it is the
use of sophisticated technological tools and computer algorithms to
rapidly trade securities. HFT uses proprietary trading strategies
carried out by computers to move in and out of positions in seconds
or fractions of a second.
\item \label{enu:Algorithmic Trading}\href{https://en.wikipedia.org/wiki/Algorithmic_trading}{Algorithmic Trading, Wikipedia Link}
Algorithmic trading is a method of executing a large order (too large
to fill all at once) using automated pre-programmed trading instructions
accounting for variables such as time, price, and volume to send small
slices of the order (child orders) out to the market over time. They
were developed so that traders do not need to constantly watch a stock
and repeatedly send those slices out manually.
\item \label{enu:Portfolio Manager}\href{https://en.wikipedia.org/wiki/Portfolio_manager}{Portfolio Manager, Wikipedia Link}
A Portfolio Manager is a professional responsible for making investment
decisions and carrying out investment activities on behalf of vested
individuals or institutions. The investors invest their money into
the portfolio manager's investment policy for future fund growth such
as a retirement fund, endowment fund, education fund and other purposes
(Bodie, Kane \& Marcus 2011). Portfolio managers work with a team
of analysts and researchers, and are responsible for establishing
an investment strategy, selecting appropriate investments and allocating
each investment properly towards an investment fund or asset management
vehicle.
\item \label{enu:Universal Computing Machine}\href{https://en.wikipedia.org/wiki/Universal_Turing_machine}{Universal Computing Machine, Wikipedia Link}
In computer science, a universal Turing machine (UTM) is a Turing
machine (Minsky 1967; End-note \ref{Turing-Machine}) that can simulate
an arbitrary Turing machine on arbitrary input. The universal machine
essentially achieves this by reading both the description of the machine
to be simulated as well as the input thereof from its own tape.
\item \label{Turing-Machine}\href{https://en.wikipedia.org/wiki/Turing_machine}{Turing Machine, Wikipedia Link}
A Turing machine is a mathematical model of computation that defines
an abstract machine, which manipulates symbols on a strip of tape
according to a table of rules. Despite the model's simplicity, given
any computer algorithm, a Turing machine capable of simulating that
algorithm's logic can be constructed (Sipser 2006).
\item \label{enu:Computer}\href{https://en.wikipedia.org/wiki/Computer}{Computer, Wikipedia Link}
A computer is a device that can be instructed to carry out sequences
of arithmetic or logical operations automatically via computer programming.
\item \label{enu: Mac or Macintosh}\href{https://en.wikipedia.org/wiki/Macintosh}{MAC or Macintosh, Wikipedia Link}
The Macintosh (pronounced as MAK-in-tosh; branded as Mac since 1998)
is a family of personal computers designed, manufactured, and sold
by Apple Inc. since January 1984.
\item \label{enu:Personal Computer}\href{https://en.wikipedia.org/wiki/Personal_computer}{Personal Computer, Wikipedia Link}
A personal computer (PC) is a multi-purpose computer whose size, capabilities,
and price make it feasible for individual use.
\item \label{enu:MAC vs MPC}\href{https://en.wikipedia.org/wiki/Apple_Computer,_Inc._v._Microsoft_Corp.}{MAC vs MPC, Wikipedia Link}
Apple Computer, Inc. v. Microsoft Corporation, was a copyright infringement
lawsuit in 1994 in which Apple Computer, Inc. (now Apple Inc.) sought
to prevent Microsoft and Hewlett-Packard from using visual graphical
user interface (GUI) elements that were similar to those in Apple's
Lisa and Macintosh operating systems. Mac vs PC also refers to the
rivalry between the two companies to dominate the personal computer
market.
\item \label{enu:History Computing}\href{https://en.wikipedia.org/wiki/History_of_computing}{History Computing, Wikipedia Link}
The history of computing is longer than the history of computing hardware
and modern computing technology and includes the history of methods
intended for pen and paper or for chalk and slate, with or without
the aid of tables.
\item \label{enu:Computing Platform}\href{https://en.wikipedia.org/wiki/Computing_platform}{Computing Platform, Wikipedia Link}
A computing platform or digital platform is the environment in which
a piece of software is executed. It may be the hardware or the operating
system (OS), even a web browser and associated application programming
interfaces, or other underlying software, as long as the program code
is executed with it.
\item \label{enu:Cloud Computing}\href{https://en.wikipedia.org/wiki/Cloud_computing}{Cloud Computing, Wikipedia Link}
Cloud computing is shared pools of configurable computer system resources
and higher-level services that can be rapidly provisioned with minimal
management effort, often over the Internet. Cloud computing relies
on sharing of resources to achieve coherence and economies of scale,
similar to a public utility. 
\item \label{enu:Quantum Computing}\href{https://en.wikipedia.org/wiki/Quantum_computing}{Quantum Computing, Wikipedia Link}
Quantum computing is computing using quantum-mechanical phenomena,
such as superposition and entanglement. A quantum computer is a device
that performs quantum computing. Such a computer is different from
binary digital electronic computers based on transistors. Whereas
common digital computing requires that the data be encoded into binary
digits (bits), each of which is always in one of two definite states
(0 or 1), quantum computation uses quantum bits or qubits, which can
be in superpositions of states.
\item \label{enu:The-Turing-test}\href{https://en.wikipedia.org/wiki/Turing_test}{Turing Test, Wikipedia Link}
The test was introduced by Turing in his 1950 paper, \textquotedbl Computing
Machinery and Intelligence\textquotedbl , while working at the University
of Manchester (Turing, 1950; p. 460). It opens with the words: \textquotedbl I
propose to consider the question, 'Can machines think?'\textquotedbl{}
Because \textquotedbl thinking\textquotedbl{} is difficult to define,
Turing chooses to \textquotedbl replace the question by another,
which is closely related to it and is expressed in relatively unambiguous
words.\textquotedbl{} Turing's new question is: \textquotedbl Are
there imaginable digital computers which would do well in the imitation
game?\textquotedbl{} This question, Turing believed, is one that can
actually be answered. In the remainder of the paper, he argued against
all the major objections to the proposition that \textquotedbl machines
can think\textquotedbl . Since Turing first introduced his test,
it has proven to be both highly influential and widely criticised,
and it has become an important concept in the philosophy of artificial
intelligence (Russell \& Norvig 2016).
\item \label{enu:The-Wolf-Wall-Street}\href{https://en.wikipedia.org/wiki/The_Wolf_of_Wall_Street_(2013_film)}{The Wolf of Wall Street (2013 Film), Wikipedia Link}
The Wolf of Wall Street is a 2013 American biographical black comedy
crime film directed by Martin Scorsese and written by Terence Winter,
based on the memoir of the same name by Jordan Belfort. It recounts
Belfort's perspective on his career as a stockbroker in New York City
and how his firm Stratton Oakmont engaged in rampant corruption and
fraud on Wall Street that ultimately led to his downfall.
\item \label{enu:Chinese room thought experiment}\href{https://en.wikipedia.org/wiki/Chinese_room\#Chinese_room_thought_experiment}{Searle's Chinese room thought experiment, Wikipedia Link}
The Chinese room argument holds that a program cannot give a computer
a \textquotedbl mind\textquotedbl , \textquotedbl understanding\textquotedbl{}
or \textquotedbl consciousness\textquotedbl , regardless of how
intelligently or human-like the program may make the computer behave.
The argument was first presented by philosopher John Searle in his
paper, \textquotedbl Minds, Brains, and Programs\textquotedbl ,
published in Behavioral and Brain Sciences in 1980. It has been widely
discussed in the years since. The centerpiece of the argument is a
thought experiment known as the Chinese room. 
\item \label{enu:Bass-Model-Diffusion}\href{https://en.wikipedia.org/wiki/Bass_diffusion_model}{Bass Model of Diffusion, Wikipedia Link}
The Bass Model or Bass Diffusion Model was developed by Frank Bass.
It consists of a simple differential equation that describes the process
of how new products get adopted in a population. The model presents
a rationale of how current adopters and potential adopters of a new
product interact. The basic premise of the model is that adopters
can be classified as innovators or as imitators and the speed and
timing of adoption depends on their degree of innovativeness and the
degree of imitation among adopters. The Bass model has been widely
used in forecasting, especially new products' sales forecasting and
technology forecasting. Mathematically, the basic Bass diffusion is
a Riccati equation (End-note \ref{enu:Ricatti-Equation}) with constant
coefficients.
\item \label{enu:Ricatti-Equation} \href{https://en.wikipedia.org/wiki/Riccati_equation}{Riccati Equation, Wikipedia Link}
In mathematics, a Riccati equation in the narrowest sense is any first-order
ordinary differential equation that is quadratic in the unknown function.
In other words, it is an equation of the form 
\[
y'(x)=q_{0}(x)+q_{1}(x)\,y(x)+q_{2}(x)\,y^{2}(x)
\]
 where $q_{0}(x)\neq0$ and $q_{2}(x)\neq0$. If $q_{0}(x)=0$ the
equation reduces to a Bernoulli equation, while if $q_{2}(x)=0$ the
equation becomes a first order linear ordinary differential equation.
The equation is named after Jacopo Riccati (1676–1754) (see Riccati
1724).
\end{doublespace}
\end{enumerate}
\begin{doublespace}

\section{References}
\end{doublespace}
\begin{enumerate}
\begin{doublespace}
\item Aldridge, I. (2013). High-frequency trading: a practical guide to
algorithmic strategies and trading systems (Vol. 604). John Wiley
\& Sons.
\item Amir, Y., Ben-Ishay, E., Levner, D., Ittah, S., Abu-Horowitz, A.,
\& Bachelet, I. (2014). Universal computing by DNA origami robots
in a living animal. Nature nanotechnology, 9(5), 353.
\item Angel, J. J., \& McCabe, D. (2013). Fairness in financial markets:
The case of high frequency trading. Journal of Business Ethics, 112(4),
585-595.
\item Au, K., Chan, F., Wang, D., \& Vertinsky, I. (2003). Mood in foreign
exchange trading: Cognitive processes and performance. Organizational
Behavior and Human Decision Processes, 91(2), 322-338.
\item Bacidore, J. M., Boquist, J. A., Milbourn, T. T., \& Thakor, A. V.
(1997). The search for the best financial performance measure. Financial
Analysts Journal, 53(3), 11-20.
\item Bahrammirzaee, A. (2010). A comparative survey of artificial intelligence
applications in finance: artificial neural networks, expert system
and hybrid intelligent systems. Neural Computing and Applications,
19(8), 1165-1195.
\item Banerjee, A. V. (1992). A simple model of herd behavior. The Quarterly
Journal of Economics, 107(3), 797-817. 
\item Banerjee, A. V. (1993). The economics of rumours. The Review of Economic
Studies, 60(2), 309-327.
\item Barberis, N., \& Thaler, R. (2003). A survey of behavioral finance.
Handbook of the Economics of Finance, 1, 1053-1128.
\item Barney, J. B. (1995). Looking inside for competitive advantage. The
Academy of Management Executive, 9(4), 49-61.
\item Bass, F. M. (1969). A new product growth for model consumer durables.
Management science, 15(5), 215-227.
\item Belfort, J. (2011). The wolf of wall street. Hachette UK.
\item Bell, R. M., \& Cover, T. M. (1980). Competitive optimality of logarithmic
investment. Mathematics of Operations Research, 5(2), 161-166.
\item Berlyne, D. E. (1954). A theory of human curiosity. British Journal
of Psychology, 45, 180-191. 
\item Berlyne, D. E. (1966). Curiosity and exploration. Science, 153. 25-33.
\item Bhattacharyya, A. (1943). On a Measure of Divergence Between Two Statistical
Populations Defined by their Probability Distributions, Bull. Calcutta
Math. Soc., 35, pp. 99-110. 26. 
\item Bhattacharyya, A. (1946). On a measure of divergence between two multinomial
populations. Sankhy\={ }a: The Indian Journal of Statistics, 401-406.
\item Biais, B., \& Woolley, P. (2011). High frequency trading. Manuscript,
Toulouse University, IDEI.
\item Bodie, Z., \& Taggart, R. A. (1978). Future investment opportunities
and the value of the call provision on a bond. The Journal of Finance,
33(4), 1187-1200.
\item Bodie, Z., Kane, A., \& Marcus, A. J. (2011). Investment and portfolio
management. McGraw-Hill Irwin.
\item Brogaard, J., Hendershott, T., \& Riordan, R. (2014). High-frequency
trading and price discovery. The Review of Financial Studies, 27(8),
2267-2306.
\item Brogaard, J., Hendershott, T., \& Riordan, R. (2017). High frequency
trading and the 2008 short-sale ban. Journal of Financial Economics,
124(1), 22-42.
\item Brown, G. W., \& Cliff, M. T. (2005). Investor sentiment and asset
valuation. The Journal of Business, 78(2), 405-440.
\item Budish, E., Cramton, P., \& Shim, J. (2015). The high-frequency trading
arms race: Frequent batch auctions as a market design response. The
Quarterly Journal of Economics, 130(4), 1547-1621.
\item Bughin, J. (2016). Big data, Big bang?. Journal of Big Data, 3(1),
2.
\item Burges, C. J. (2009). Dimension reduction: A guided tour. Machine
Learning, 2(4), 275-365.
\item Carroll, L. (1865). (2012 Reprint) Alice's adventures in wonderland.
Random House, Penguin Random House, Manhattan, New York.
\item Carroll, L. (1871). (2009 Reprint) Through the looking glass: And
what Alice found there. Random House, Penguin Random House, Manhattan,
New York. 
\item Cartea, A., Jaimungal, S., \& Ricci, J. (2014). Buy low, sell high:
A high frequency trading perspective. SIAM Journal on Financial Mathematics,
5(1), 415-444.
\item Cartea, Á., \& Jaimungal, S. (2015). Risk metrics and fine tuning
of high‐frequency trading strategies. Mathematical Finance, 25(3),
576-611.
\item Cartea, Á., Jaimungal, S., \& Kinzebulatov, D. (2016). Algorithmic
trading with learning. International Journal of Theoretical and Applied
Finance, 19(04), 1650028.
\item Cameron, J., \& Wisher, W. (1991). Terminator 2: judgment day (Vol.
2). USA.
\item Carlton, J., \& Annotations-Kawasaki, G. (1997). Apple: The inside
story of intrigue, egomania, and business blunders. Random House Inc.
\item Ceruzzi, P. E. (2003). A history of modern computing. MIT press.
\item Chaboud, A. P., Chiquoine, B., Hjalmarsson, E., \& Vega, C. (2014).
Rise of the machines: Algorithmic trading in the foreign exchange
market. The Journal of Finance, 69(5), 2045-2084.
\item Chen, T. (2013). Do investors herd in global stock markets?. Journal
of Behavioral Finance, 14(3), 230-239.
\item Chiang, T. C., \& Zheng, D. (2010). An empirical analysis of herd
behavior in global stock markets. Journal of Banking \& Finance, 34(8),
1911-1921.
\item Chirinko, R. S., \& Wilson, D. J. (2008). State investment tax incentives:
A zero-sum game?. Journal of Public Economics, 92(12), 2362-2384.
\item Christie, W. G., \& Huang, R. D. (1995). Following the pied piper:
Do individual returns herd around the market?. Financial Analysts
Journal, 51(4), 31-37.
\item Chung, S. L., Hung, C. H., \& Yeh, C. Y. (2012). When does investor
sentiment predict stock returns?. Journal of Empirical Finance, 19(2),
217-240.
\item Cooper, R., Davis, M., \& Van Vliet, B. (2016). The mysterious ethics
of high-frequency trading. Business Ethics Quarterly, 26(1), 1-22.
\item Cramér, H. (2016). Mathematical methods of statistics (PMS-9) (Vol.
9). Princeton university press.
\item Dasgupta, S., \& Gupta, A. (1999). An elementary proof of the Johnson-Lindenstrauss
lemma. Inter- national Computer Science Institute, Technical Report,
99-006.
\item Davis, M. (2011). The universal computer: The road from Leibniz to
Turing. AK Peters/CRC Press.
\item Davis, M., Kumiega, A., \& Van Vliet, B. (2013). Ethics, finance,
and automation: a preliminary survey of problems in high frequency
trading. Science and engineering ethics, 19(3), 851-874.
\item De Prado, M. M. L., \& Peijan, A. (2004). Measuring loss potential
of hedge fund strategies. The Journal of Alternative Investments,
7(1), 7-31.
\item Derpanis, K. G. (2008). The Bhattacharyya Measure. Mendeley Computer,
1(4), 1990-1992.
\item Elton, E. J., Gruber, M. J., Brown, S. J., \& Goetzmann, W. N. (2009).
Modern portfolio theory and investment analysis. John Wiley \& Sons.
\item Evans, J. L. (2004). Wealthy investor attitudes, Expectations, and
Behaviors toward risk and return. The Journal of Wealth Management,
7(1), 12-18.
\item Fama, E. F. (1970). Efficient capital markets: A review of theory
and empirical work. The journal of Finance, 25(2), 383-417.
\item Ferguson, R., \& Leistikow, D. (1998). Search for the best financial
performance measure: basics are better. Financial Analysts Journal,
54(1), 81-85.
\item Fodor, I. K. (2002). A survey of dimension reduction techniques. Technical
Report UCRL-ID-148494, Lawrence Livermore National Laboratory.
\item Formaini, R. (2001). Free markets on film: Hollywood and capitalism.
Journal of Private Enterprise, 16(2), 123.
\item Foster, F. D., \& Viswanathan, S. (1990). A theory of the interday
variations in volume, variance, and trading costs in securities markets.
The Review of Financial Studies, 3(4), 593-624.
\item Fox, C. R., \& Levav, J. (2000). Familiarity bias and belief reversal
in relative likelihood judgment. Organizational Behavior and Human
Decision Processes, 82(2), 268-292.
\item Frankl, P., \& Maehara, H. (1988). The Johnson-Lindenstrauss lemma
and the sphericity of some graphs. Journal of Combinatorial Theory,
Series B, 44(3), 355-362.
\item Frankl, P., \& Maehara, H. (1990). Some geometric applications of
the beta distribution. Annals of the Institute of Statistical Mathematics,
42(3), 463-474. 
\item Freiberger, P., \& Swaine, M. (1999). Fire in the Valley: the making
of the personal computer. McGraw-Hill Professional.
\item French, R. M. (2000). The Turing Test: the first 50 years. Trends
in cognitive sciences, 4(3), 115-122.
\item Fung, W., \& Hsieh, D. A. (2004). Hedge fund benchmarks: A risk-based
approach. Financial Analysts Journal, 60(5), 65-80.
\item Gibbons, R. (1992). A primer in game theory. Harvester Wheatsheaf.
\item Giot, P., \& Laurent, S. (2003). Value‐at‐risk for long and short
trading positions. Journal of Applied Econometrics, 18(6), 641-663.
\item Gupta, M., \& George, J. F. (2016). Toward the development of a big
data analytics capability. Information \& Management, 53(8), 1049-1064.
\item Hagen, U. (1991). Challenge for the Actor. Simon and Schuster.
\item Hamadène, S. (2006). Mixed zero-sum stochastic differential game and
American game options. SIAM Journal on Control and Optimization, 45(2),
496-518.
\item Hand, D. J. (1998). Data mining: Statistics and more?. The American
Statistician, 52(2), 112-118.
\item Hawley, D. D., Johnson, J. D., \& Raina, D. (1990). Artificial neural
systems: A new tool for financial decision-making. Financial Analysts
Journal, 46(6), 63-72.
\item Haykin, S. S. (2004). Neural networks: A comprehensive foundation.
\item Hill, J. M. (2006). Alpha as a net zero-sum game. The Journal of Portfolio
Management, 32(4), 24-32.
\item Holt, J. (1995). How Children Learn. Classics in Child Development.
\item Holton, G. A. (2004). Defining risk. Financial Analysts Journal, 60(6),
19-25.
\item Huberman, G. (2001). Familiarity breeds investment. The Review of
Financial Studies, 14(3), 659-680.
\item Johnson, W. B., \& Lindenstrauss, J. (1984). Extensions of Lipschitz
mappings into a Hilbert space. Contemporary mathematics, 26(189-206),
1.
\item Kashyap, R. (2014). The Circle of Investment. International Journal
of Economics and Finance, 6(5), 244-263. 
\item Kashyap, R. (2016). Notes on Uncertainty, Unintended Consequences
and Everything Else. Working Paper.
\item Kashyap, R. (2017). Artificial Intelligence: A Child's Play. Working
Paper.
\item Kirilenko, A., Kyle, A. S., Samadi, M., \& Tuzun, T. (2017). The Flash
Crash: High‐frequency trading in an electronic market. The Journal
of Finance, 72(3), 967-998.
\item Kryzanowski, L., Galler, M., \& Wright, D. W. (1993). Using artificial
neural networks to pick stocks. Financial Analysts Journal, 49(4),
21-27.
\item Kumiega, A., \& Van Vliet, B. E. (2012). Automated finance: The assumptions
and behavioral aspects of algorithmic trading. Journal of Behavioral
Finance, 13(1), 51-55.
\item Kyle, A. S. (1985). Continuous auctions and insider trading. Econometrica:
Journal of the Econometric Society, 1315-1335.
\item Laraki, R., \& Solan, E. (2005). The value of zero-sum stopping games
in continuous time. SIAM Journal on Control and Optimization, 43(5),
1913-1922.
\item Lee, K. Y., \& Bretschneider, T. R. (2012). Separability Measures
of Target Classes for Polarimetric Synthetic Aperture Radar Imagery.
Asian Journal of Geoinformatics, 12(2).
\item Leuenberger, M. N., \& Loss, D. (2001). Quantum computing in molecular
magnets. Nature, 410(6830), 789. 
\item Li, K., Cooper, R., \& Van Vliet, B. (2017). How Does High-Frequency
Trading Affect Low-Frequency Trading?. Journal of Behavioral Finance,
1-14.
\item Loewenstein, G. (1994). The psychology of curiosity: A review and
reinterpretation. Psychological bulletin, 116(1), 75. 
\item Loewy, E. H. (1998). Curiosity, imagination, compassion, science and
ethics: Do curiosity and imagination serve a central function?. Health
Care Analysis, 6(4), 286-294.
\item Manes, S., \& Andrews, P. (1993). Gates: How Microsoft’s mogul reinvented
an industry-and made himself the richest man in America. Simon \&
Schuster. 
\item Menkveld, A. J. (2013). High frequency trading and the new market
makers. Journal of Financial Markets, 16(4), 712-740.
\item Minsky, M. L. (1967). Computation: finite and infinite machines. Prentice-Hall,
Inc.
\item Muñoz Torrecillas, M. J., Yalamova, R., \& McKelvey, B. (2016). Identifying
the Transition from Efficient-Market to Herding Behavior: Using a
Method from Econophysics. Journal of Behavioral Finance, 17(2), 157-182.
\item Nash, J. F. (1950). Equilibrium points in n-person games. Proceedings
of the national academy of sciences, 36(1), 48-49.
\item Norstad, J. (1999). The normal and lognormal distributions.
\item Osborne, M. J., \& Rubinstein, A. (1994). A course in game theory.
MIT press.
\item Paddock, J. R., Terranova, S., \& Giles, L. (2001). SASB goes Hollywood:
Teaching personality theories through movies. Teaching of psychology,
28(2), 117-121.
\item Perrier, J. Y., Sipper, M., \& Zahnd, J. (1996). Toward a viable,
self-reproducing universal computer. Physica D: Nonlinear Phenomena,
97(4), 335-352.
\item Popper, K. R. (2002). The poverty of historicism. Psychology Press.
\item Porter, L. (2014). Benedict Cumberbatch, Transition Completed: Films,
Fame, Fans. Andrews UK Limited.
\item Preston, J., \& Bishop, M. J. (2002). Views into the Chinese room:
New essays on Searle and artificial intelligence. OUP.
\item Provost, F., \& Fawcett, T. (2013). Data science and its relationship
to big data and data-driven decision making. Big data, 1(1), 51-59.
\item Rao, C. R. (1973). Linear statistical inference and its applications
(Vol. 2, pp. 263-270). New York: Wiley.
\item Reilly, F. K., \& Brown, K. C. (2002). Investment analysis and portfolio
management.
\item Reio Jr, T. G., Petrosko, J. M., Wiswell, A. K., \& Thongsukmag, J.
(2006). The measurement and conceptualization of curiosity. The Journal
of Genetic Psychology, 167(2), 117-135.
\item Riccati, J. (1724). Animadversiones in aequationes differentiales
secundi gradus. Actorum Eruditorum Supplementa, 8(1724), 66-73.
\item Russell, S. J., \& Norvig, P. (2016). Artificial Intelligence: A Modern
Approach. Prentice-Hall, Engle- wood Cliffs, 25, 27.
\item Savani, R. (2012). High-frequency trading: The faster, the better?.
IEEE Intelligent Systems, 27(4), 70-73.
\item Schwartz, R. H., \& Smith, D. E. (1988). Hallucinogenic mushrooms.
Clinical pediatrics, 27(2), 70-73.
\item Searle, J. R. (1980). Minds, brains, and programs. Behavioral and
brain sciences, 3(3), 417-424.
\item Seiler, M. J., Seiler, V. L., Harrison, D. M., \& Lane, M. A. (2013).
Familiarity bias and perceived future home price movements. Journal
of Behavioral Finance, 14(1), 9-24.
\item Shu, H. C., \& Chang, J. H. (2015). Investor Sentiment and Financial
Market Volatility. Journal of Behavioral Finance, 16(3), 206-219.
\item Simon, H. A. (1962). The Architecture of Complexity. Proceedings of
the American Philosophical Society, 106(6), 467-482.
\item Sipser, M. (2006). Introduction to the Theory of Computation (Vol.
2). Boston: Thomson Course Technology.
\item Sorzano, C. O. S., Vargas, J., \& Montano, A. P. (2014). A survey
of dimensionality reduction techniques. arXiv preprint arXiv:1403.2877.
\item Taleb, N. N. (2007). The black swan: the impact of the highly improbable.
NY: Random House. 
\item Tambe, P. (2014). Big data investment, skills, and firm value. Management
Science, 60(6), 1452-1469.
\item Thompson, K. F., Gokler, C., Lloyd, S., \& Shor, P. W. (2016). Time
independent universal computing with spin chains: quantum plinko machine.
New Journal of Physics, 18(7), 073044.
\item Treleaven, P., Galas, M., \& Lalchand, V. (2013). Algorithmic trading
review. Communications of the ACM, 56(11), 76-85.
\item Turing, A. M. (1950). Computing machinery and intelligence. Mind,
59(236), 433-460.
\item Turnbull, S. M. (1987). Swaps: a zero sum game?. Financial Management,
16(1), 15-21.
\item Von Neumann, J., \& Morgenstern, O. (1953). Theory of games and economic
behavior. Princeton university press.
\item Vuorenmaa, T. A. (2013). The good, the bad, and the ugly of automated
high-frequency trading. The Journal of Trading, 8(1), 58-74.
\item Wang, M., Keller, C., \& Siegrist, M. (2011). The less You know, the
more You are afraid of—A survey on risk perceptions of investment
products. Journal of Behavioral Finance, 12(1), 9-19.
\item Williams, M. R. (1997). A history of computing technology. IEEE Computer
Society Press. 
\item Wong, B. K., \& Selvi, Y. (1998). Neural network applications in finance:
a review and analysis of literature (1990–1996). Information \& Management,
34(3), 129-139.
\item Wonglimpiyarat, J. (2012). Technology strategies and standard competition—Comparative
innovation cases of Apple and Microsoft. The Journal of High Technology
Management Research, 23(2), 90-102
\item Wooldridge, M., \& Jennings, N. R. (1995). Intelligent agents: Theory
and practice. The knowledge engineering review, 10(2), 115-152.
\item You, J. (2015). Beyond the Turing test. Science, 347(6218), 116-116.
\item Young, H. P. (2009). Learning by trial and error. Games and economic
behavior, 65(2), 626-643.
\item Zhang, Q., Cheng, L., \& Boutaba, R. (2010). Cloud computing: state-of-the-art
and research challenges. Journal of internet services and applications,
1(1), 7-18. 
\end{doublespace}
\end{enumerate}

\end{document}